\documentclass[doublecol]{epl2} 

\newcommand{\ket}[1]{|#1\rangle}
\newcommand{\braket}[2]{\langle #1|#2\rangle}
\usepackage{amsmath}
\title{Aharonov-Bohm-like effect for light propagating in nematics with disclinations}
\shorttitle{Aharonov-Bohm-like effect for photons in nematics with disclinations} 

\author{A. M. de M. Carvalho\inst{1} \and C. S\'atiro\inst{2} \and F. Moraes\inst{2}}
\shortauthor{A. M. de M. Carvalho \etal}

\institute{                    
  \inst{1} Departamento de F\'{\i}sica, Universidade Estadual de Feira de Santana,
BR116-Norte, Km 3,
44031-460, Feira de Santana, BA, Brazil\\
  \inst{2} Departamento de F\'{\i}sica, CCEN,  Universidade Federal 
da Para\'{\i}ba, Caixa Postal 5008, 58051-970 Jo\~ao Pessoa, PB,
Brazil
}

\abstract{Using a geometric approach for the propagation of light in anisotropic media, we investigate what effect the director field of disclinations may have on the polarization state of light. Parallel transport around the defect, of the spinor describing the polarization,  indicates the acquisition  of a topological phase, in analogy with the Aharonov-Bohm effect.}

\begin{document}

\maketitle

\section{Introduction}
In a recent article \cite{sat} two of us (CS and FM) studied the propagation of light in nematics with topological defects using a geometric approach due to Joets and Ribotta \cite{joe}. The model associates Fermat's principle of geometrical optics to geodesics in a hypothetic non-Euclidean space. This way, an effective geometry can be found which may be interpreted as ``the cause'' of the bending of the light paths, just as in general relativity. An important issue concerning this approach is the possibility of using the mighty resources of differential geometry to calculate physical properties of such system. Indeed, in this Letter we show that spinors,  which describe the polarization of a photon, when parallelly tranported around a loop encircling a disclination, acquire a topological phase analogous to the Aharonov-Bohm effect \cite{aha}.

\section{Effective geometry for disclinations in nematics}
We study wedge disclinations of generic strength $k$, whose director configurations, in the plane $x-y$, are given by \cite{kle}
\begin{equation}
\varphi(\theta)=k\theta+c , \label{phi}
\end{equation}
where $\varphi$ is the angle  between the molecular axis and the $x$-axis, $\theta$ is the angular polar coordinate and $c=\varphi(0)$. Selected director configurations can be seen on Figure 11.4 of \cite{kle}. We assume the disclinations are straight and lie along the $z$-axis and the light rays propagate in the $x$-$y$ plane so, effectively, we have a two-dimensional problem.

Eq. (\ref{phi}) implies that $\varphi(2\pi)-\varphi(0)=2\pi k$. Since the nematic molecules are symmetric under a rotation of $\pi$ radians around an axis perpendicular to the molecular axis, this result must be $n\pi$ where $n$ is a positive or negative integer. This means that $k$ can only take integer or half-integer values.

In ref. \cite{sat} we found that the effective geometry perceived by light traveling in the neighborhood of a disclination, in a nematic liquid crystal, is described by the  line element
\begin{eqnarray}
& ds^2 & = \left\{\cos^{2}\xi+\alpha^2 \sin^{2}\xi\right\}d\rho^{2} \nonumber\\
& + & \left\{\sin^{2}\xi+\alpha^2 \cos^{2}\xi\right\}\rho^{2}d\theta^{2} \nonumber\\
& - & \left\{2(\alpha^2-1)\sin\xi\cos\xi\right\}\rho d\rho d\theta ,\label{metric}
 \nonumber\\
& & 
\end{eqnarray}
where $\xi$ depends on the azimuthal angle $\theta$ as
\begin{equation}
\xi(\theta)=(k-1)\theta+c
\end{equation}
and where $\alpha=n_e/n_o$ and $n_{o}$ and $n_{e}$ are the ordinary and extraordinary refractive indices, respectively. The radial coordinate $r$, used in ref. \cite{sat}, has been rescaled to $\rho=n_o r$.  

The above metric is associated to a curvature scalar given by
\begin{equation}
R=\frac{2k(k-1)(1-\alpha^{2})}{\alpha^{2}\rho^{2}}\cos(2\xi). \label{scalar}
\end{equation}
In fact, as we show below, this result is valid only for $\rho\neq 0$ since the idealized coreless defects will have a $\delta$-function curvature singularity at $\rho=0$.

\section{Parallel transport of a vector in the effective geometry}
Although our main interest in this work is the parallel transport of spinors, this section is important because it introduces the contribution of the singularity at the $z$-axis to the curvature, which justifies the main result of this Letter. The parallel transport of a vector in a loop involving curved space is equivalent to a local rotation of the vector. The rotation angle contains precious information on the nature of the encircled curved space. We use this information to infer the contribution of the singularity at $\rho=0$ to the curvature. To do this we use the powerful and concise calculus of differential forms \cite{dar}. 

We introduce an appropriate dual $1$-form basis (coframe), defined in terms of  local \textit{zweibein} fields by $e^{a}=e^{a}_{\mu}dx^{\mu}$, where
\begin{subequations}
\begin{eqnarray}
        e^{1}&=& \left( \cos\xi d\rho+\sin\xi \rho d\theta\right) \mbox{,}  \label{e1}\\
        e^{2}&=& \alpha \left( \sin\xi d\rho-\cos\xi \rho d\theta\right) \mbox{.}  \label{e2}
\end{eqnarray}
\end{subequations}
We introduce an affine spin connection 1-form $\omega^{a}_{b}$ and define the torsion 2-form and the curvature 2-form, respectively, by
\begin{subequations}
\begin{eqnarray}
T^{a}&=&\frac{1}{2}T^{a}_{bc}\;e^{b}\wedge e^{c}=de^{a}+\omega^{a}_{b} \wedge e^{b}\mbox{,} \\
R^{a}_{b}&=&\frac{1}{2}R^{a}_{bcd}\; e^{c}\wedge e^{d}.
\end{eqnarray}
\end{subequations}
These are the Maurer-Cartan structure equations. Since we already know that the effective geometry is Riemannian \cite{sat} we make use of the torsion-free condition for this geometry, which implies that the first of the Maurer-Cartan equations becomes
\begin{equation}
de^{a}+\omega^{a}_{\mu} \wedge e^{b}=0.
\end{equation}
From the 1-form basis we can determine the non-null  connection 1-forms:
\begin{eqnarray}
\omega^{1}_{2}&=&-\omega^{2}_{1}=k\left[\frac{1-\alpha^2}{\alpha}\sin\xi\cos\xi\frac{d\rho}{\rho}\right.\nonumber\\ 
   &+&\left.\left(\frac{1}{\alpha}+\frac{\alpha^{2}-1}{\alpha}\cos^{2}\xi\right)d\theta \right]   \mbox{.}
\end{eqnarray}
The  connection 1-forms  transform in the same way as the gauge potential of a non-Abelian gauge theory, which means that any two elements of the group do not commute.

This result takes us to the following spin connection for the effective metric (\ref{metric})
\begin{eqnarray}
\Gamma_{\theta} =\left( \begin{array}{cccc}
0 & B  \\ 
-B & 0 
\end{array}\right) ,
\end{eqnarray}
and
\begin{eqnarray}
\Gamma_{r} =
\left( \begin{array}{cccc}
A & 0 \\ 
0 & -A 
\end{array}\right) ,
\end{eqnarray}
where
\begin{subequations}
\begin{eqnarray}
A&=& \frac{1-\alpha^2}{\alpha}\kappa\;\sin\xi\cos\xi\   \mbox{,} \\
B&=& \frac{\kappa}{\alpha}+\kappa\frac{\alpha^{2}-1}{\alpha}\cos^{2}\xi.
\end{eqnarray}
\end{subequations}
The spin connection $\Gamma_{\theta}$ corresponds to orbits with the radial coordinate $\rho$ constant and  $\Gamma_{\rho}$  corresponds to curves with the $\theta$ coordinate constant. The holonomy matrix $U(\gamma)$ associated with the parallel transport of vectors around a closed curve $\gamma$  is defined by
\begin{equation}
U(\gamma)= {\cal P}\exp \left[ -\oint_{\gamma}dx^{\mu} \Gamma_{\mu} \right],\label{holo}
\end{equation}
where ${\cal P}$ is the ordering operator. The set of all holonomy matrices forms the holonomy group which is defined as the group of linear transformations of the tangent space $T_{p}M$ induced by parallel transport around loops based at a point $p$. This group possesses information regarding to the curvature of the effective geometry.

We are interested in analyzing circular orbits. In this case the holonomy possess only the azimuthal contribution to the spin connection and is written as
\begin{equation}
U(\gamma_{\theta})={\cal P}\exp \left(-\oint \Gamma_{\theta}d\theta \right), \label{U}
\end{equation}
which results in
\begin{eqnarray}
U(\gamma_{\theta}) =
\left( \begin{array}{cccc}
\cos\lambda & -\sin\lambda \\ 
\sin\lambda & \cos\lambda 
\end{array}\right) \label{matrix},
\end{eqnarray}
where
\begin{eqnarray}
\lambda&=&\left(\frac{\pi}{\alpha}+\alpha\pi\right)k \nonumber\\        &+&\frac{\alpha^{2}-1}{\alpha}\left\{\frac{\sin\left[4\pi(k-1)+2c\right]-\sin2c}{4(k-1)}\right\}k \label{lambda}
\end{eqnarray}
The matrix $U(\gamma_{\theta})$  gives the effective rotation of the parallelly transported vector but it includes a $2\pi$ rotation due to the loop. Therefore $\chi=2\pi -\lambda$ is the angle between the final and initial versions of the vector.

Eq. (\ref{lambda}) gets greatly simplified when the allowed values of $k$ (integer or half-integer) are taken into account. For $k=1$ it is easily seen that
\begin{equation}
\chi(k=1)=2\pi(1-\alpha) \label{chik1}
\end{equation}
and for the remaining integer or half-integer $k$
\begin{equation}
\chi(k\neq1)=2\pi-\pi\left[\alpha+\frac{1}{\alpha}\right]k .\label{chik2}
\end{equation}

Notice that, when $k=1$ and $c=0$ the metric (\ref{metric}) reduces to that of a cosmological object, the cosmic string, as already reported in ref. \cite{sat2} and, accordingly, the outcome of our calculations reduces to the know  result for the cosmic string \cite{val}, where $\chi=2\pi(1-\alpha)$. This, at first sight, may seem incompatible with the null curvature obtained when one makes $k=1$ in equation (\ref{scalar}). In fact, as it is well known in the cosmic string lore there is a $\delta$-function curvature singularity at $\rho=0$ which is responsible for the change in the direction of the transported vector. As we show below this also happens for the generic $k$ case.

\begin{figure}[!h]
\begin{center}
\includegraphics[height=1.3in]{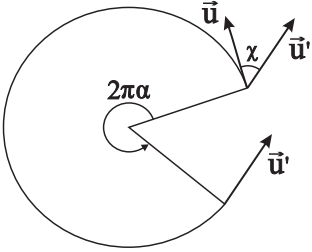} 
\caption{Parallel transport of a tangent vector around a circle concentric with the tip of a cone. The angle between the initial and final versions of the vector is $\chi=2\pi(1-\alpha)$.}
\label{cone}
\end{center}
\end{figure}

Before analysing the more general case we recall that for $k=1$ and $c=0$ the geometry is that of a cone \cite{sat2}. Fig. \ref{cone} clearly shows the effect of the parallel transport of a vector alongside a circle in the Euclidean plane where a wedge of angle $2\pi(1-\alpha)$ was removed in order to make the cone by identification of the lose edges. The resulting angle between the initial and final vectors is $\chi=2\pi(1-\alpha)$, in agreement with eq. (\ref{chik1}).

Now, the above results have to be compatible with the Gauss-Bonnet theorem \cite{man} which states that, for a regular Riemannian surface, 
\begin{equation}
 \int_{Q} K dA = 2\pi-\int_{\partial Q}k_{g}ds, \label{gaussbonnet}
\end{equation}
where $K$ is the Gaussian curvature, $\partial Q$ is the boundary of the singly-connected domain $Q$ and $k_{g}$ is the geodesic curvature of the line $\partial Q$. Since our interest is circular orbits, we take $Q$ to be a disc of radius $\rho$ centered at $\rho=0$ and its circumference to be $\partial Q$.

There is a very nice  relationship between the Gauss-Bonnet theorem and the element of the holonomy group (\ref{matrix}) giving the angle of local rotation of a vector after parallel transport in a loop \cite{vic},
\begin{equation}
\chi=2\pi-\int_{\partial Q}k_{g}ds=\int_{Q} K dA,
\end{equation}
which we explicitly verify below.

For a curve in a surface with line element $ds^2=Edu^2+2Fdudv+Gdv^2$, the geodesic curvature is \cite{man}
\begin{eqnarray}
 k_g&=&\sqrt{EG-F^2}\left[-\Gamma_{11}^2 \dot{u}^3 +\Gamma_{22}^1 \dot{v}^3-(2\Gamma_{12}^2 -\Gamma_{11}^1)\dot{u}^2\dot{v} \right. \nonumber \\
    &+&\left. (2\Gamma_{12}^1 -\Gamma_{22}^2)\dot{u}\dot{v}^2 +\ddot{u}\dot{v}-\ddot{v}\dot{u}\right],
\end{eqnarray}
where $\Gamma_{ij}^{k}$ are the Christoffel symbols of second kind and the derivatives are taken with respect to an affine parameter along the curve, which we choose to be $s$ itself. We choose $u=\theta$ and $v=\rho$.
The metric given by eq. (\ref{metric}) yields $\sqrt{EG-F^2}=\alpha\rho$ and, since $\partial Q$ is a circle $\rho=const$, we have 
\begin{equation}
ds=\sqrt{\sin^{2}\xi+\alpha^2 \cos^{2}\xi}\rho d\theta ,
\end{equation}
from (\ref{metric}). This gives immediately 
\begin{equation}
\dot{\theta}=\rho^{-1}(\sin^{2}\xi+\alpha^2 \cos^{2}\xi)^{-1/2}.
\end{equation} 
The geodesic curvature reduces then to
\begin{equation}
k_g =-\alpha\rho\Gamma_{\theta\theta}^r \dot{\theta}^3 .
\end{equation}
Now, using the metric (\ref{metric}) we obtain
\begin{eqnarray}
\Gamma_{\theta\theta}^r=-\frac{\rho}{\alpha^2}\left\{\alpha^2(1-k)+k[1-(1-\alpha^2)\cos^2\xi]^2\right\}.
\end{eqnarray}
Finally,
\begin{eqnarray}
\int_{\partial Q}k_g ds&=&\int_{0}^{2\pi}\left\{\frac{(1-k)\alpha}{\sin^{2}\xi+\alpha^2 \cos^{2}\xi}\right. \nonumber\\
&+&\left.\frac{k}{\alpha}[\sin^{2}\xi+\alpha^2 \cos^{2}\xi]\right\}d\theta, 
\end{eqnarray}
with $\xi(\theta)=(k-1)\theta+c$. These integrals are easily solved to
\begin{equation}
\int_{\partial Q}k_g ds=2\pi\alpha 
\end{equation}
for $k=1$ and
\begin{equation}
\int_{\partial Q}k_g ds=\pi\left[\alpha+\frac{1}{\alpha}\right]k
\end{equation}
for $k\neq 1$ integer or half-integer.

Since we are working in two-dimensional space, the curvature scalar is related to the Gaussian curvature by $K=R/2$. Using this, we substitute eq. (\ref{scalar}) into the left hand side of eq. (\ref{gaussbonnet}), with $dA=\alpha\rho d\rho d\theta$, which is easily obtained by taking the wedge product between $e^1$ and $e^2$ (eqs. (\ref{e1}) and (\ref{e2})). For $k=1$ we have that $K=R/2=0$ but $\chi=2\pi\alpha$ indicates that there should be a $\delta$-function contribution to $K$. Therefore, in order that the Gauss-Bonnet theorem is satisfied, 
\begin{equation}
K=\left(\frac{1-\alpha}{\alpha}\right)\frac{\delta(\rho)}{\rho} \label{gauss1}
\end{equation}
for $k=1$, in agreement with ref. \cite{sok}.

Now, for $k\neq 1$ the integral $\int_{q} K dA=0$ again as it can be easily verified taking $K=R/2$ and using eq. (\ref{scalar}). Again, it means that eq. (\ref{scalar}) is incomplete since $\chi=2\pi-\pi\left[\alpha+\frac{1}{\alpha}\right]k$. Therefore 
\begin{eqnarray}
K&=&\frac{1}{\alpha}\left[1-\left(\alpha+\frac{1}{\alpha}\right)\frac{k}{2}\right]\frac{\delta(\rho)}{\rho}\nonumber\\
 &+&\frac{k(k-1)(\alpha^{2}-1)}{\alpha^{2}\rho^{2}}\cos\left\{2\left[(k-1)\theta+c\right]\right\} \label{gauss2}
\end{eqnarray} 
for $k\neq 1$ integer or half-integer. This means that, even though for $\rho\neq 0$ there is a $\rho$- and $\theta$-dependent curvature associated to the disclination, the average of this curvature term in a disk centered in the origin is null. The singularity there is the sole responsible for the parallel transport effects. In real systems, the curvature $\delta$-function singularity in the core of the defect will be smoothed out. But still, one can think of a localized flux of curvature accross a very small area, the disclination core cross secion. So, even though the light ray does not crosses the core of the disclination, its spin angular momentum will ``feel'' the curvature flux, in a clear analogy with the Aharonov-Bohm effect. This justifies the next section on the parallel transport of a spinor.

\section{Parallel transport of a spinor in the effective geometry}
An important question that emerges when we study the parallel transport of vectors is what happens with more complex fields when they also undergo a parallel transport. Vectors in a curved background have their orientation changed after a complete loop if there is a nonvanishing average curvature in the region surrounded by the loop. Spinors, as well, have their components changed if the same basis is used. This can be interpreted as a rotation in spinor space or the acquisition of a topological phase like Berry's \cite{ber}.

Motivated by the spin angular momentum of the light, or polarization, we are interested in the study of the parallel transport of a spinor in the effective geometry associated to the disclinations.  The spinorial connection is
\begin{equation}
\label{holonomy2}
\Gamma_{\mu}(x)=-\frac{1}{4}\omega^{\alpha}_{\nu \mu}\gamma_{\alpha}\gamma^{\nu},
\end{equation}
where $\gamma_{\alpha}$ are the flat-space Dirac matrices and $\omega^{\alpha}_{\nu \mu}$ are the coefficients of the spin connection. The holonomy matrix is given, as in the parallel transport of vectors, by eq. (\ref{holo}) but in terms of the spinorial connection $\Gamma_{\mu}(x)$ instead of the spin connection.

For circular orbits we find the following expression for the spinorial connection in terms of the Dirac matrices
\begin{eqnarray}
\Gamma_{\theta}(x)&=&-\frac{1}{4}\omega^{1}_{2\theta}\left(\gamma_{1}\gamma^{2}-\gamma_{2}\gamma^{1}
\right)\nonumber \\
&=&\frac{1}{2} \omega^{1}_{2\theta}\, i\,\sigma^{z},
\end{eqnarray}
where we have used the following representation  for the $\gamma$ matrices in terms of the Pauli matrices: $\gamma^{1}=i\sigma^{y}$ and $\gamma^{2}=-i\sigma^{x}$. Again, the phase associated with the above spinorial connection is given by eq. (\ref{U}). Integrating and expanding the exponential in eq. (\ref{U}) we obtain
\begin{equation}
U(\gamma_{\theta})=\cos(\chi/2)I+i\sin(\chi/2)\sigma^{z},
\end{equation}
where $I$ is the identity matrix, $\sigma^{z}$ the Pauli matrix and $\chi$ is given either by eq. (\ref{chik1}) or eq. (\ref{chik2}). Therefore, when a two-component spinor is parallelly transported in a closed orbit around the curvature flux given by eq. (\ref{gauss1}) or eq. (\ref{gauss2}) it undergoes a phase change given by 
\begin{equation}
\ket{\psi'}=U(\gamma_{\theta})\ket{\psi}= \left(\begin{matrix}
 e^{i\chi/2} & 0 \\
 0 & e^{-i\chi/2}
                                          \end{matrix}\right)
\ket{\psi}, \label{qua}
\end{equation}
with phase angles corresponding, respectively, to eq.(\ref{chik1}) and eq. (\ref{chik2}).

\section{Discussion}

\begin{figure}[!h]
\begin{center}
\includegraphics[height=1.3in]{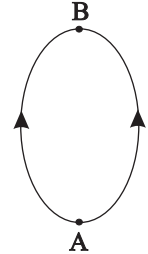} 
\caption{Integration paths for eq. (\ref{U}).}
\label{cam}
\end{center}
\end{figure}

The phase (\ref{qua}) was obtained for a circular path around the line defect. But this is a topological effect and therefore the loop can be deformed without changing the result as far as the defect core is kept inside the loop. Moreover, since the line integral $\oint = \int_A^B + \int_B^A=\int_A^B - \int_A^B$, following different paths (see Fig. \ref{cam}), eq. (\ref{U}) can be split into two parts, each integral to be done on a separate path between points $A$ and $B$. 
This way, the result (\ref{qua}) gives the polarization phase difference between two light beams travelling in opposite sides of the defect. 

Considering a beam of linearly polarized light travelling past a disclination, if the input state is
\begin{equation}
\ket{\psi}=\frac{1}{\sqrt{2}}\left(\ket{+}+\ket{-}\right),
\end{equation}
the output state, obtained from eq. (\ref{qua}), is
\begin{equation}
\ket{\psi'}=\frac{1}{\sqrt{2}}\left[\exp(i\chi/2)\ket{+}+\exp(-i\chi/2)\ket{-}\right].
\end{equation}
The internal product between these two states is
\begin{equation}
\vert\braket{\psi}{\psi'}\vert^2=\frac{1}{4}\left[\exp(i\chi/2)+\exp(-i\chi/2)\right]^2=\cos^2(\chi/2). \label{rot}
\end{equation}
Then, according to Malus's law, the polarization plane of the light rotates by a relative angle $\chi/2$. That is, if the original beam is split in two, each of them travelling on either side of the defect, the relative rotation of the polarization plane is given by the angle $\chi/2$ when they are rejoined. Similar results apply to circularly and elliptically polarized beams, as can be easily verified.

The rotation of the polarization in eq. (\ref{rot}) is analogous to what is observed when a linearly polarized laser beam propagates down a nonplanar optical fiber \cite{ross}. This effect has been interpreted as a manifestation of Berry's phase \cite{chi,tom}. Like the original Aharonov-Bohm effect \cite{aha}, where a flux of magnetic field provokes a phase shift in a split electron beam, the manifestation of Berry's phase can be associated to a flux of a gauge field in parameter space \cite{sim}.  In our case, the  topological defect is associated to a flux of curvature in ordinary space. Since only the curvature associated to the defect core contributes, the phase is topological and should not depend on the local geometry. On the other hand, the defect is a particular configuration of nematic molecules. If one or a few of them have their orientation disturbed it is no longer the same defect. Then the phase associated to this deformed defect may not be the same as the one obtained here. Thermal agitation and molecular reorientation due to the electric field of a polarized laser beam are then effects that may affect experimental measurements of the predicted phase shift. To minimize thermal  effects the sample should be as far below the isotropic to nematic transition temperature as possible. Since the reorientation angle is very small for laser beams with low power \cite{self,hat} the minimum possible power should be used. The closest the beam passes the core of the defect the better because  the strong ``anchoring'' of the nematic molecules near the core will offer resistance to the torque provoked by the external field. 

In order to do an experiment to measure the effect predicted here one needs disclinations in the nematic sample. These defects are spontaneously generated at the isotropic to nematic phase transition. In a recent article, Mukai and coworkers \cite{hat2} produced disclinations in a lyotropic mixture (potassium laurate, decanol and water) in a sealed planar glass cell. Although their interest was in the formation statistics of the defects (which by the way represents another analogy with cosmic strings) their procedure for generating and identifying the defects can be followed. The defects are observed in an optical microscope with the sample placed between crossed polarizers. If a binocular microscope is used, one of the oculars can be used to focus the laser beam near a chosen defect and the second polarizer rotated in order to maximize the intensity at a photodiode conveniently placed after the second polarizer. This will give the phase change in a single beam travelling past a disclination. One can try also an interference experiment with the aid of the converging lens behavior predicted for some of the disclinations \cite{sat} since they naturally force light rays following paths on opposing sides of the defect to meet. In this case the relative phase shift between the polarizations of the beams on different sides of the defect is measured. Although in the original Aharonov-Bohm effect there is no deflection of the beam, the analogy is in the fact that in both cases there is a confined field flux causing a topological phase shift. As described above, the deflection is an advantage that can help in the detection of the effect.

Diffraction of light by a disclination will certainly be affected by the effective geometry described here.  Although the more interesting cases will be those of asymmetric defects (under present study) we can infer some properties of the $k=1$ case (symmetric defect) from their cosmic analogues. In reference \cite{sat2} it was shown that the $k=1$ defect is analogous to a cosmic string in the sense that they share the same space geometry. Linet \cite{lin} and, more recently, Yamamoto and Tsunoda \cite{yam} investigated the propagation of a plane wave past a cosmic string in the limit of geometrical optics. From their work we obtain that the cylindrical diffracted wave has an order of magnitude  $\frac{A\alpha}{2\pi}$, where $A$ is the amplitude of the incident plane wave and $\alpha=\frac{n_o}{n_e}$. For cosmic strings \cite{vil}   $\alpha=1-4G\mu/c^2\approx 10^{-6}$, where $G$ is the gravitational constant and $\mu$ is the mass density of the string, which for grand unification strings $\mu\approx 10^{22}g/cm$. This indicates that, while the diffraction effect may be very difficult to observe for cosmic strings, it may be observable for disclinations in nematic liquid crystals which have $\alpha\approx 10^0$. 

A recent experiment that is somehow related to the results presented here was done by Marrucci and coworkers \cite{angular}. Injecting circularly polarized light in q plates, which are in fact cross sections of the defects discussed here, they observed the conversion of the spin angular momentum into orbital angular momentum. The incident Gaussian beam is therefore converted to a helical mode preserving total angular momentum in this process. Their setup is exactly orthogonal to ours but nevertheless a change in the helicity is observed although not in the sense of our work. In our case there is only a rotation of the polarization without conversion of spin to angular momentum therefore keeping a Gaussian beam in its original mode. 

A number of experiments have been carried out on the manifestation of Berry's phase for the photon polarization. Among them is the observation of the rotation  of linearly polarized light propagating down a helically wound, single-mode optical fiber \cite{ross,tom}. We have seen above that the polarization of light propagating in the neighborhood of disclinations is affected in a similar way by the director field of the defect. Using a geometric approach we showed that this polarization change is due to a flux of curvature associated to the core of the disclination. This is clearly a manifestation of Berry's phase and analogous to the Aharonov-Bohm effect in the sense that the spinorial connection $\Gamma_{\mu}$ has the same role as the vector potential $A_{\mu}$ in provoking the phase change. Similar effects were predicted for disclinations in elastic solids \cite{azevedo} and for femionic quasiparticles in superfluids \cite{gar}. 

\acknowledgments
We thank CNPq and CAPES (PROCAD program) and FINEP/FAPESQ (PRONEX program). We are also indebted to J. Schaum, C. Furtado, H. Mukai and P. R. G. Fernandes for important comments and suggestions.


\begin{thebibliography}{0}
\bibitem{sat} 
  \Name{S\'atiro C. \and Moraes F.}
  \REVIEW{Eur. Phys. J. E}{20}{2006}{173}.
\bibitem{joe}
  \Name{Joets A. \and Ribotta R.}
  \REVIEW{Optics Commun.}{107}{1994}{200}.
\bibitem{aha}
  \Name{Aharonov Y. \and Bohm D.}
  \REVIEW{Phys. Rev.}{115}{1959}{485}.
\bibitem{kle} 
  \Name{Kl\'eman M. \and Lavrentovich O. D.}
  \Book{Soft Matter Physics}
  \Publ{Springer-Verlag, New York}
  \Year{2003}.
  \Section{11.1}.
\bibitem{dar}
  \Name{Darling R. W. R.}
  \Book{Differential Forms and Connections}
  \Publ{Cambridge University Press, Cambridge}
  \Year{1999}.
\bibitem{sat2}
  \Name{S\'atiro C. \and Moraes F.}
  \REVIEW{Mod. Phys. Lett. A}{20}{2005}{2561}.
\bibitem{val}
  \Name{Bezerra V. B.}
  \REVIEW{Phys. Rev. D}{35}{1987}{2031}.
\bibitem{man}
  \Name{do Carmo M. P.}
  \Book{Riemannian Geometry}
  \Publ{Birkh\"auser, Boston}
  \Year{1992}.
\bibitem{vic}
  \Name{Vickers J. A. G.}
  \REVIEW{Class. Quantum Grav.}{4}{1987}{1}.
\bibitem{sok}
  \Name{Sokolov D. D. \and Starobinskii A. A.}
  \REVIEW{Sov. Phys. Dokl.}{22}{1977}{312}. 
\bibitem{self}
  \Name{Brugioni S. and Meucci R.}
  \REVIEW{Eur. Phys. J. D}{28}{2004}{277}.
\bibitem{hat}
  \Name{Fernandes P. R. G., Mukai H., Honda B. S. L. \and Shibli S. M.}
  \REVIEW{Liq. Cryst.}{33}{2006}{367}.
\bibitem{hat2}
  \Name{Mukai H., Fernandes P. R. G., Oliveira B. F. \and Dias G. S.}
  \REVIEW{Phys. Rev. E}{75}{2007}{061704}.
\bibitem{ber}
  \Name{Berry M. V.}
  \REVIEW{Proc. R. Soc. London}{A392}{1984}{45}. 
\bibitem{ross}
  \Name{Ross J. N.}
  \REVIEW{Opt. Quant. Electron.}{16}{1984}{455}.
\bibitem{lin}
  \Name{Linet B.}
  \REVIEW{Ann. Inst. Henri Poincar\'e}{45}{1986}{249}.
\bibitem{yam}
  \Name{Yamamoto K. \and Tsunoda K.}
  \REVIEW{Phys. Rev. D}{68}{2003}{041302}.
\bibitem{vil}
  \Name{Vilenkin A.}
  \REVIEW{Phys. Rev. D}{23}{1981}{852}. 
\bibitem{chi}
  \Name{Chiao R. Y \and Wu Y. S.}
  \REVIEW{Phys. Rev. Lett.}{57}{1986}{933}.
\bibitem{tom}
  \Name{Tomita A. \and Chiao R. Y.}
  \REVIEW{Phys. Rev. Lett.}{57}{1986}{937}. 
\bibitem{sim}
  \Name{Simon B.}
  \REVIEW{Phys. Rev. Lett.}{51}{1983}{2167}.
\bibitem{angular}
  \Name{Marrucci L., Manzo C. \and Paparo D.}
  \REVIEW{Phys. Rev. Lett.}{96}{2006}{163905}.
\bibitem{azevedo}
  \Name{Azevedo S. \and Moraes F.}
  \REVIEW{Phys. Lett. A}{246}{1998}{374}. 
\bibitem{gar}
  \Name{Garcia de Andrade L. C., de M. Carvalho A. M. \and Furtado C. }
  \REVIEW{Europhys. Lett.}{67}{2004}{538}.
\end{thebibliography}
\end{document}